\documentclass[proceedings]{JHEP3} 
\usepackage{epsf}
\usepackage{axodraw}
\usepackage{epsfig,multicol}			

\newbox\mybox
\newcommand\fverb{\setbox\mybox=\hbox\bgroup\verb}
\newcommand\fverbdo{\egroup\medskip\noindent\fbox{\unhbox\mybox}\ }
\newcommand\fverbit{\egroup\item[\fbox{\unhbox\mybox}]}

\def\lsim{\raise0.3ex\hbox{$\;<$\kern-0.75em\raise-1.1ex\hbox{$\sim\;$}}}
\def\gsim{\raise0.3ex\hbox{$\;>$\kern-0.75em\raise-1.1ex\hbox{$\sim\;$}}}

\title{CP violation in production of neutralinos in $e^+ e^-$ collisions} 
 
\author{\speaker{O. Kittel} \\
Institut f\"ur Theoretische Physik und Astrophysik, Universit\"at
W\"urzburg, 
Am Hubland, 
D-97074~W\"urzburg, Germany 
and \\Institut de F\'{\i}sica Corpuscular - C.S.I.C., 
Universitat de Val{\`e}ncia 
Edifici Instituts d'Investigaci{\'o}, 
- Apartat de Correus 22085 - 
E-46071 Val{\`e}ncia, Spain\\
	E-mail:  
\email{kittel@physik.uni-wuerzburg.de \\ 
WUE-ITP-03-023, IFIC-03-52, hep-ph/0311169}}

\conference{International Workshop on Astroparticle and High Energy Physics}

\abstract{
We propose T-odd and CP-odd odd asymmetries
in order to analyze the impact of CP violating phases
in neutralino production  and subsequent leptonic 
two-body decays.
We present numerical results of these 
asymmetries and of the cross sections
for complex parameters $\mu$, $M_1$ and $A_{\tau}$,
which may be present in the neutralino and stau sector 
of the Minimal Supersymmetric Standard Model if CP is violated.
The asymmetries arise on tree level
and thus could be large enough to be observed
at a linear $ e^+e^-$ collider in the $\sqrt{s}=500$ GeV range
with high luminosity.
We discuss the feasibility for measuring the asymmetries
by analyzing their statistical errors.
Moreover we study the beam polarization dependence of the 
asymmetries and of the cross sections and show that they both
can be enhanced considerably.
}

\begin{document}


\section{Introduction}

The only source of CP violation in the Standard Model  is
given by one phase in the Kobayashi-Maskawa matrix.
%
However, this phase alone cannot account for the observed baryon
asymmetry of the universe \cite{BGSM}, and further
sources of CP violation have to be introduced.
In the Minimal Supersymmetric Standard Model (MSSM)
several supersymmetric (SUSY) breaking 
parameters and the higgsino mass parameter $\mu$ 
can be complex.
%
%
The  phases of the SUSY parameters are restricted by the experimental 
upper limits on the electric dipole moments (EDMs) 
\cite{Masiero:xj} of electron, 
neutron and of the $^{199}$Hg and $^{205}$Tl atoms.
%
%
However, there can be strong cancellations between the
different SUSY contributions to the EDMs,
which can weaken the restrictions on the phases \cite{edmstheo}.
%
%
Independently from the EDMs, an unambiguous determination
of the values of the phases is necessary in order to clarify 
whether the SUSY phases are candidates for causing the  
baryon asymmetry of the universe.

The phases have also impact on the phenomenology of 
production and decay of SUSY particles, in particular 
at a future linear $ e^+e^-$
collider \cite{TDR}, and give rise to an important class of CP and T-odd 
observables, which involve triple products 
\cite{donoghue,oshimo,choi1}.
They allow us to define various CP and T asymmetries which are
sensitive to the different CP phases.
On the one hand, these observables are naturally large because
they are  present at tree level. 
On the other hand, they also allow a determination of
the sign of the phases, which could not be achieved if only
CP-even observables would be studied.

In this talk, we study neutralino production
(for recent studies with complex parameters and polarized beams see
\cite{choi1,kali,gudi1} )
\begin{eqnarray} \label{production}
   e^++e^-&\to&
	\tilde{\chi}^0_i+\tilde{\chi}^0_j 
\end{eqnarray}
and the subsequent leptonic two-body decay of one of the neutralinos
\begin{eqnarray} \label{decay_1}
   \tilde{\chi}^0_i&\to& \tilde{\ell} + \ell_1,  
\end{eqnarray}
and of the decay slepton
	\begin{eqnarray} \label{decay_2}
		\tilde{\ell}&\to&\tilde{\chi}^0_1+ \ell_2;\;\;\; \ell_{1,2}=
		e,\mu,\tau.
\end{eqnarray}
The triple product 
$ {\mathcal T }=( {\bf  p}_{e^-} \times 
	{\bf p}_{\ell_2})\cdot {\bf p}_{\ell_1}$
defines the T-odd asymmetry of the cross section $\sigma$
for the processes (\ref{production})-(\ref{decay_2}):
\begin{eqnarray} \label{Tasymmetry}
{\mathcal A}_{\rm T} = \frac{\sigma({\mathcal T}>0)
				 -\sigma({\mathcal T}<0)}
					{\sigma({\mathcal T}>0)+
					\sigma({\mathcal T}<0)}.
\end{eqnarray}
The dependence of  ${\mathcal A}_{\rm T}$ on 
$\varphi_{M_1}$ and $\varphi_{\mu}$, which are
the phases of the complex gaugino mass parameter $M_1$ 
and the complex higgsino  mass parameter $\mu$,
respectively, was studied 
in \cite{olaf}, and for the neutralino three-body
decays also in \cite{Barger:2001nu,oshimo,choi1,karl}.

In case the neutralino decays into a $\tau$-lepton,
$\tilde{\chi}^0_i\to\tilde{\tau}_k^{\pm} \tau^{\mp}$,
$k=1,2$,
the transverse $\tau^-$ polarization $P_2$
and the transverse $\tau^+$ polarization $\bar P_2$,
give rise to the CP-odd asymmetry 
\begin{eqnarray} \label{ACP}
{\mathcal A}_{\rm CP}=\frac{1}{2}(P_2-\bar{P}_2),
\end{eqnarray}
which is also sensitive to the phases $\varphi_{A_{\tau}}$
of the complex trilinear scalar coupling parameter
$A_{\tau}$ of the stau sector of the MSSM.
Without measuring the transverse $\tau^{\pm}$-polarizations,  a sensitivity
to $\varphi_{A_{\tau}}$ is not obtained \cite{Bartl:2003ck}, 
and one would only be sensitive
to CP violation in the production process (\ref{production}) \cite{olaf}.
The components of the $\tau^-$ polarization 
vector are  given by \cite{Renard,staupol}
\begin{eqnarray}
	{\rm P}_i &=&\frac{ {\rm Tr}(\varrho  \, \sigma_i  )}
	{ {\rm Tr}(\varrho)},
\end{eqnarray}
with $\varrho$ the hermitean spin density matrix of the $\tau^-$
and $\sigma_i$  the Pauli matrices.
The $\tau^-$ polarization ${\bf P}=(P_1,P_2,P_3)$ 
is defined in a coordinate 
system in which  $P_3$ is the longitudinal polarization,
and $P_1$  is the transverse polarization in the plane formed by 
${\bf p}_{e^-}$ and ${\bf p}_{\tau}$.
The T-odd component $P_2$ is the polarization perpendicular to
${\bf p}_{\tau}$ and ${\bf p}_{e^-}$ and is proportional to the 
triple product
${\bf s}_{\tau}^2\cdot({\bf p}_{\tau}\times {\bf p}_{e^-})$ 
\cite{staupol}, where ${\bf s}_{\tau}^2$ is the $\tau$ spin basis vector.
The dependence of  ${\mathcal A}_{\rm CP}$ on 
$\varphi_{M_1}$,  $\varphi_{\mu}$  and $\varphi_{A_{\tau}}$
was studied in \cite{staupol,Choi:2003pq}.
Also the beam polarization dependence of 
${\mathcal A}_{\rm T}$, ${\mathcal A}_{\rm CP}$  \cite{POWER} 
and of the cross sections \cite{POWER,gudi2} was studied.
\FIGURE[h]{
\setlength{\unitlength}{1cm}
\begin{picture}(15,6.)(-2,.5)
	 \put(1.,4.7){${\bf p}_{\chi_j}$}
	 \put(3.4,6.1){${\bf p}_{e^- }$}
	\put(3.3,2.3){${\bf p}_{e^+}$}
	\put(5.0,4.7){${\bf p}_{\chi_i}$}
	\put(7.4,6.0){${\bf p}_{\ell_1}$}
	 \put(6.2,3.){${\bf p}_{\tilde{ \ell} }$}
	\put(9.1,2.4){${\bf p}_{\ell_2}$}
	 \put(6.,1.4){${\bf p}_{\chi_1}$}
\end{picture}
\scalebox{1.9}{
\begin{picture}(2.5,0)(1.2,-.5)
\ArrowLine(40,50)(0,50)
\Vertex(40,50){2}
\ArrowLine(55,80)(40,50)
\ArrowLine(25,20)(40,50)
\ArrowLine(40,50)(80,50)
\ArrowLine(80,50)(110,75)
\DashLine(80,50)(100,20){4}
\Vertex(80,50){2}
\ArrowLine(100,20)(125,15)
\ArrowLine(100,20)(85,0)
\Vertex(100,20){2}
\end{picture}}
\caption{\label{shematic picture}
          Schematic picture of the neutralino production
          and decay process.}
}

\section{Numerical results
	\label{Numerical results}}

We present numerical results for 
$e^+e^-\to\tilde\chi^0_1 \tilde\chi^0_2$
with the subsequent leptonic decay of $ \tilde\chi^0_2$
for a linear collider with $\sqrt{s}=500$ GeV and longitudinally
polarized beams.
For ${\mathcal A}_{\rm T}$, Eq.~(\ref{Tasymmetry}), 
we study the neutralino 
decay  into the right selectron and right smuon, 
$\tilde \chi^0_2\to\tilde\ell_R\ell_1$, $\ell=e,\mu$ and
for ${\mathcal A}_{\rm CP}$, Eq.~(\ref{ACP}),
that  into the lightest scalar tau,
$\tilde \chi^0_2\to\tilde\tau_1\tau$.
For a schematic picture of the production and decay
process, see Fig.~\ref{shematic picture}.
We study the dependence of the asymmetries 
and the cross sections on the parameters
$\mu = |\mu| \, e^{ i\,\varphi_{\mu}}$, 
$M_1 = |M_1| \, e^{ i\,\varphi_{M_1}}$, 
$A_{\tau} = |A_{\tau}| \, e^{ i\,\varphi_{A_{\tau}}}$
and on the beam polarizations $P_{e^-}$ and $P_{e^+}$.
We assume $|M_1|=5/3 M_2\tan^2\theta_W $ and use the
renormalization group equations \cite{hall} for the 
selectron and smuon masses,
$m_{\tilde\ell_R}^2 = m_0^2 +0.23 M_2^2
-m_Z^2\cos 2 \beta \sin^2 \theta_W$ with $m_0=100$ GeV.
The interaction Lagrangians and details on stau mixing can
be found in \cite{olaf}.

For the calculation of the neutralino
width and branching ratios we neglect three-body decays 
and include the following two-body decays
\begin{eqnarray}
	\tilde\chi^0_2 &\to& \tilde\tau_{m}\tau,~
\tilde\ell_{R,L}\ell,~ 
\tilde\chi^0_1 Z,~
\tilde\chi^{\mp}_n W^{\pm},~
\tilde\chi^0_1 H_1^0,
\quad \ell=e,\mu, \quad m,n=1,2,
\end{eqnarray}
with $H_1^0$ being the lightest neutral Higgs boson.
The Higgs parameter is chosen $m_{A}=1000$~GeV.    
The decays  $\tilde\chi^0_2 \to \tilde\chi^{\pm}_n H^{\mp}$,
with $ H^{\mp}$ being the charged Higgs bosons,
and the decays $\tilde\chi^0_2 \to \tilde\chi^0_1~H_{2,3}^0$,
with $ H_{2,3}^0$ being the neutral Higgs bosons,
are thus ruled out in our scenarios.
For the  branching ratios of the sleptons we take
${\rm BR}(\tilde\ell_R\to\tilde\chi^0_1\ell)=1$, for 
$\ell=e,\mu$.

\subsection{Asymmetry ${\mathcal A}_{\rm T}$
	\label{asymmetryAT}}

In Fig.~\ref{plot_2}a we show contour lines of the 
asymmetry  ${\mathcal A}_{\rm T}$ in the $|\mu|$--$M_2$ plane
for $\varphi_{M_1}=0.1\pi $ and $\varphi_{\mu}=0$. 
It is remarkable that ${\mathcal A}_{\rm T}$ can be close to 6\%, 
even for the small value of $\varphi_{M_1}=0.1\pi $
and for $\varphi_{\mu}=0$.
A small value of the phases, especially $|\varphi_{\mu}| \lsim \pi/10 $
\cite{Barger:2001nu}, is suggested by constraints on
electron and neutron EDMs.
In Fig.~\ref{plot_2}b we show the cross section $\sigma=
\sigma(e^+e^-\to\tilde\chi^0_1\tilde\chi^0_2 ) \times
{\rm BR}(\tilde \chi^0_2\to\tilde\ell_R\ell_1)\times
{\rm BR}(\tilde\ell_R\to\tilde\chi^0_1\ell_2)$,
summed over both signs of charge of $ \ell=e,\mu$,
with BR($ \tilde\ell_R \to \tilde\chi^0_1\ell_2$) = 1.
Both  ${\mathcal A}_{\rm T}$ and
$\sigma$ also depend sensitively on the polarizations of the
$e^+$ and $e^-$-beams \cite{POWER}.
The choice $(P_{e^-},P_{e^+})=(0.8,-0.6)$
enhances the contributions
of the right slepton exchange in the neutralino production,
Eq.~(\ref{production}), and reduces those of left slepton exchange. 
The contributions of right and left slepton exchange enter
${\mathcal A}_{\rm T}$ with  opposite sign, which accounts for
the sign change of ${\mathcal A}_{\rm T}$ in Fig.~\ref{plot_2}a.

%
\FIGURE[t]{
\setlength{\unitlength}{1cm}
\begin{picture}(10,8)(2.65,0)
   \put(0,0){\includegraphics{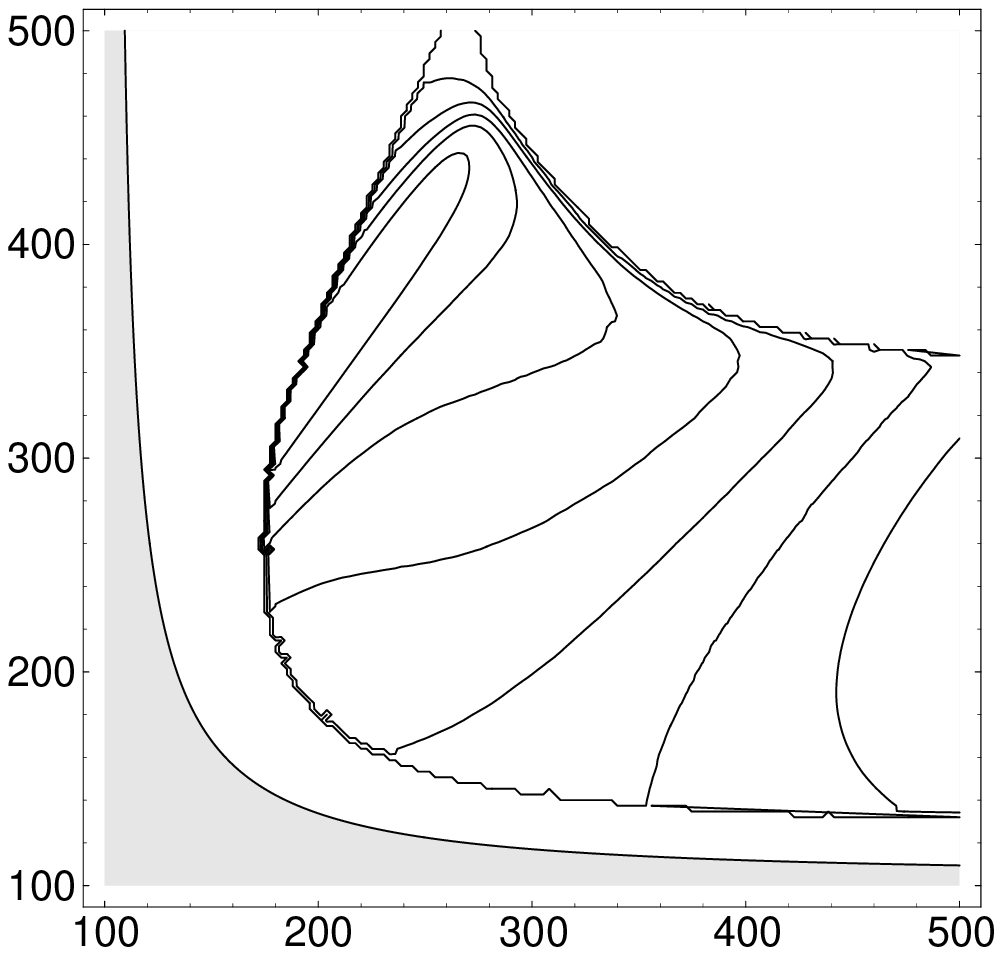}}
	\put(3.,7.4){\fbox{${\mathcal A}_{\rm T}$ in \% }}
	\put(5.5,-0.3){$|\mu|$~/GeV}
	\put(0,7.4){$M_2$~/GeV }
	\put(2.8,5.3){\footnotesize$ 6 $}
	\put(3.2,5.1){\footnotesize$ 4 $}
	\put(3.8,4.5){\footnotesize$3 $}
	\put(4.4,3.8){\footnotesize$2 $}
	\put(4.8,3.5){\footnotesize$1$}
	\put(5.9,2.9){\footnotesize$0 $}
  \put(6.2,2.6){\footnotesize$-1 $}
  	\put(5.5,6){\begin{picture}(1,1)(0,0)
			\CArc(0,0)(7,0,380)
			\Text(0,0)[c]{{\footnotesize A}}
	\end{picture}}
			\put(1.7,5.5){\begin{picture}(1,1)(0,0)
			\CArc(0,0)(7,0,380)
			\Text(0,0)[c]{{\footnotesize B}}
		\end{picture}}
\put(0.5,-.3){Fig.~\ref{plot_2}a}
	\put(8,0){\includegraphics{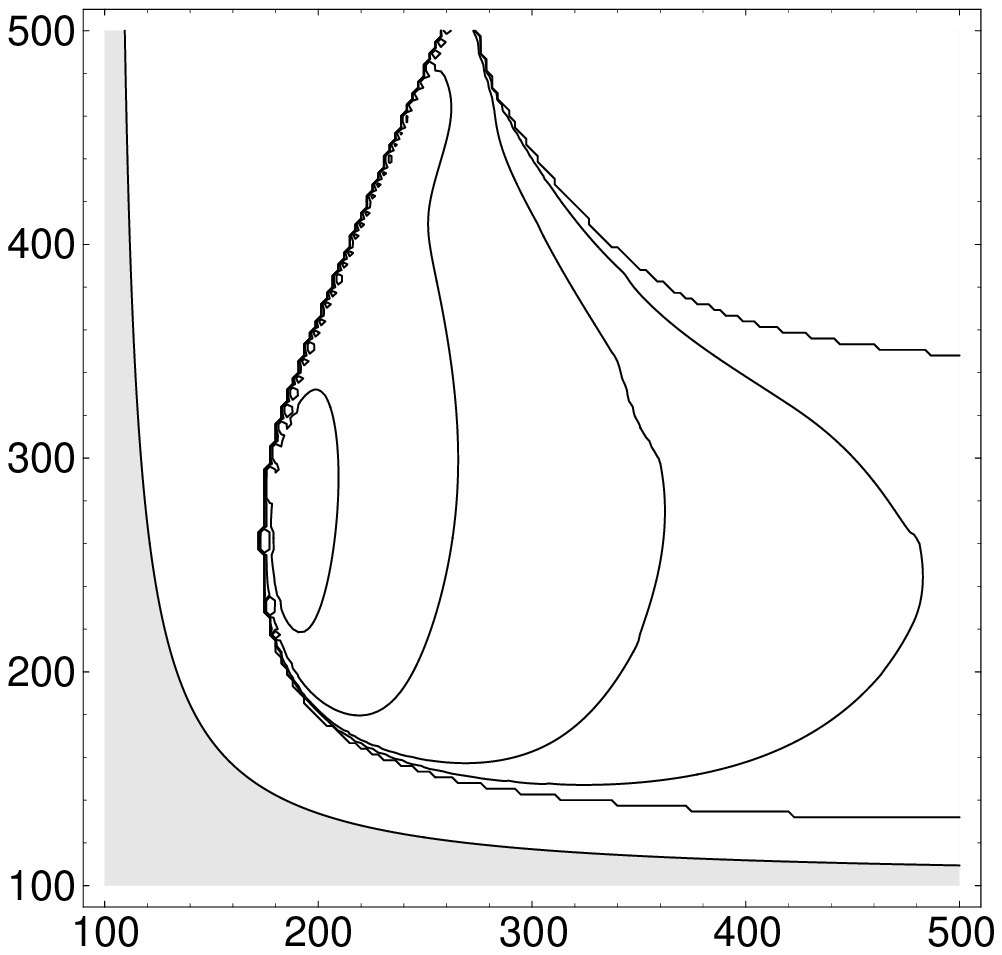}}
	\put(10.1,7.4){\fbox{$\sigma(e^+\,e^- \to\tilde{\chi}^0_1 
				\tilde{\chi}^0_1 \ell_1 \ell_2 )$ in fb}}
	\put(13.5,-.3){$|\mu|$~/GeV}
	\put(8,7.4){$M_2$~/GeV }
	\put(9.9,3.2){\footnotesize$120$}
	\put(10.7,3.0){$40$}
	\put(12.2,2.5){$4 $}
	\put(13.7,2.2){$0.4 $}
	  	\put(13.5,6){\begin{picture}(1,1)(0,0)
			\CArc(0,0)(7,0,380)
			\Text(0,0)[c]{{\footnotesize A}}
	\end{picture}}
			\put(9.7,5.5){\begin{picture}(1,1)(0,0)
			\CArc(0,0)(7,0,380)
			\Text(0,0)[c]{{\footnotesize B}}
		\end{picture}}
	\put(8.5,-.3){Fig.~\ref{plot_2}b}
\end{picture}
\vspace*{.5cm}
\caption{
	Contour lines of 
	the asymmetry ${\mathcal A}_{\rm T}$ (\ref{plot_2}a) and 
	$\sigma=\sigma(e^+e^-\to\tilde\chi^0_1\tilde\chi^0_2 ) \times
	{\rm BR}(\tilde \chi^0_2\to\tilde\ell_R\ell_1)\times
	{\rm BR}(\tilde\ell_R\to\tilde\chi^0_1\ell_2)$ (\ref{plot_2}b),
	with BR($ \tilde\ell_R \to\tilde\chi^0_1\ell_2$) = 1,
	in the $|\mu|$--$M_2$ plane for $\varphi_{M_1}=0.1\pi $, 
	$\varphi_{\mu}=0$, taking  $\tan \beta=10$, $m_0=100$ GeV,
	$A_{\tau}=-250$ GeV, $\sqrt{s}=500$ GeV 
	and $(P_{e^-},P_{e^+})=(0.8,-0.6)$.
	The area A (B) is kinematically forbidden since
	$m_{\tilde\chi^0_1}+m_{\tilde\chi^0_2}>\sqrt{s}$
	$(m_{\tilde\ell_R}>m_{\tilde\chi^0_2})$.
	The gray  area is excluded by $m_{\tilde\chi_1^{\pm}}<104 $ GeV.
\label{plot_2}}
}

\begin{figure}[h]
\setlength{\unitlength}{1cm}
			\begin{minipage}{0.45\textwidth}
 \begin{picture}(10,8)(.5,0)
	\put(0,0){\includegraphics{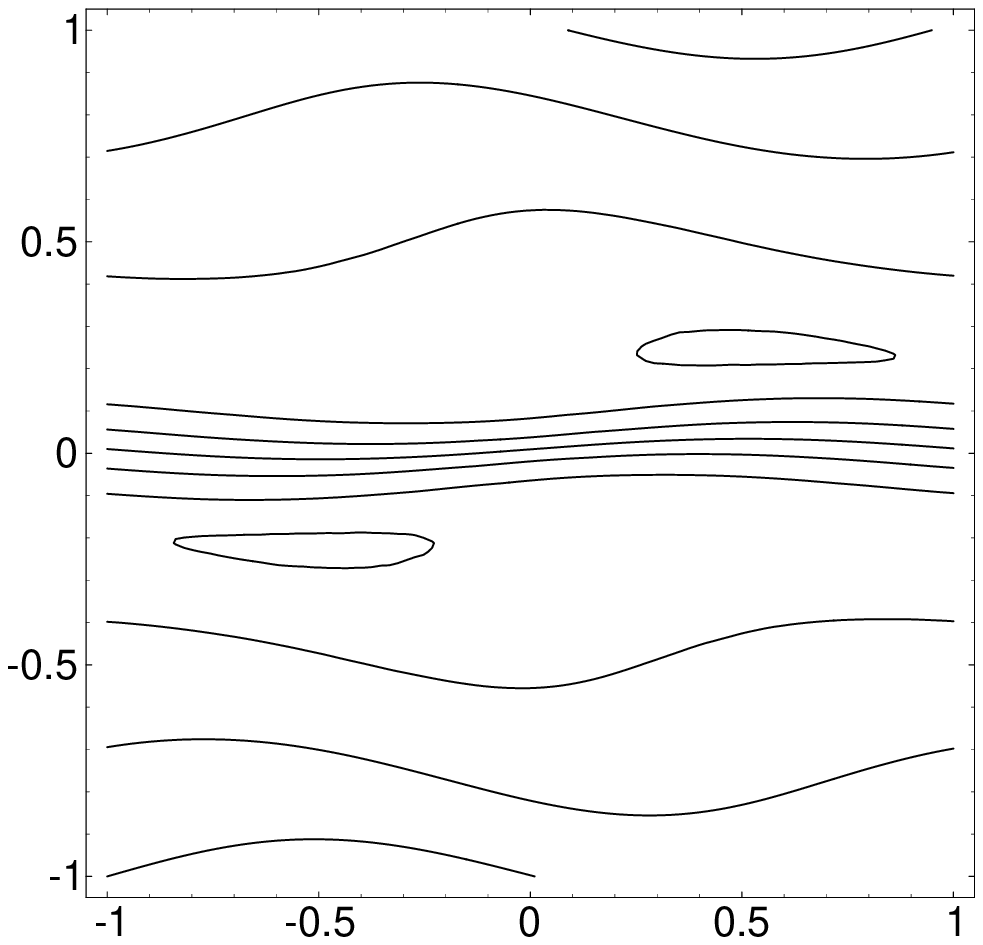}}
	\put(3.3,7.5){\fbox{${\mathcal A}_{\rm T}$ in \% }}
	\put(6.5,-.3){$\varphi_{\mu}/\pi$}
	\put(0.3,7.3){$ \varphi_{M_1}/\pi$ }
	\put(5.5,6.6){\footnotesize$0$}
	\put(3.1,6.1){\footnotesize$5$}
	\put(3.7,5.1){\footnotesize$8$}
	\put(5.2,4.4){\footnotesize$8.9$}
	\put(4.,4.2){\footnotesize$8$}
		\put(2.2,3){\footnotesize$-8.9$}
		\put(4.3,3.2){\footnotesize$-8$}
		\put(4.,2.3){\footnotesize$-8$}
		\put(4.4,1.3){\footnotesize$-5$}
		\put(2.3,0.7){\footnotesize$0$}
 \end{picture}
\vspace*{.1cm}
\caption{Contour lines of the asymmetry ${\mathcal A}_{\rm T}$
in the $\varphi_{\mu}$--$\varphi_{M_1}$ plane
for $M_2=400$ GeV and $|\mu|=240$ GeV, taking
$\tan \beta=10$, $m_0=100$ GeV,
$\sqrt{s}=500$ GeV and $(P_{e^-},P_{e^+})=(0.8,-0.6)$.
\label{plot_3}}
\end{minipage}
\hspace*{0.1cm}
	\begin{minipage}{0.45\textwidth}
 \begin{picture}(10,8)(.3,0)
	\put(0,0){\includegraphics{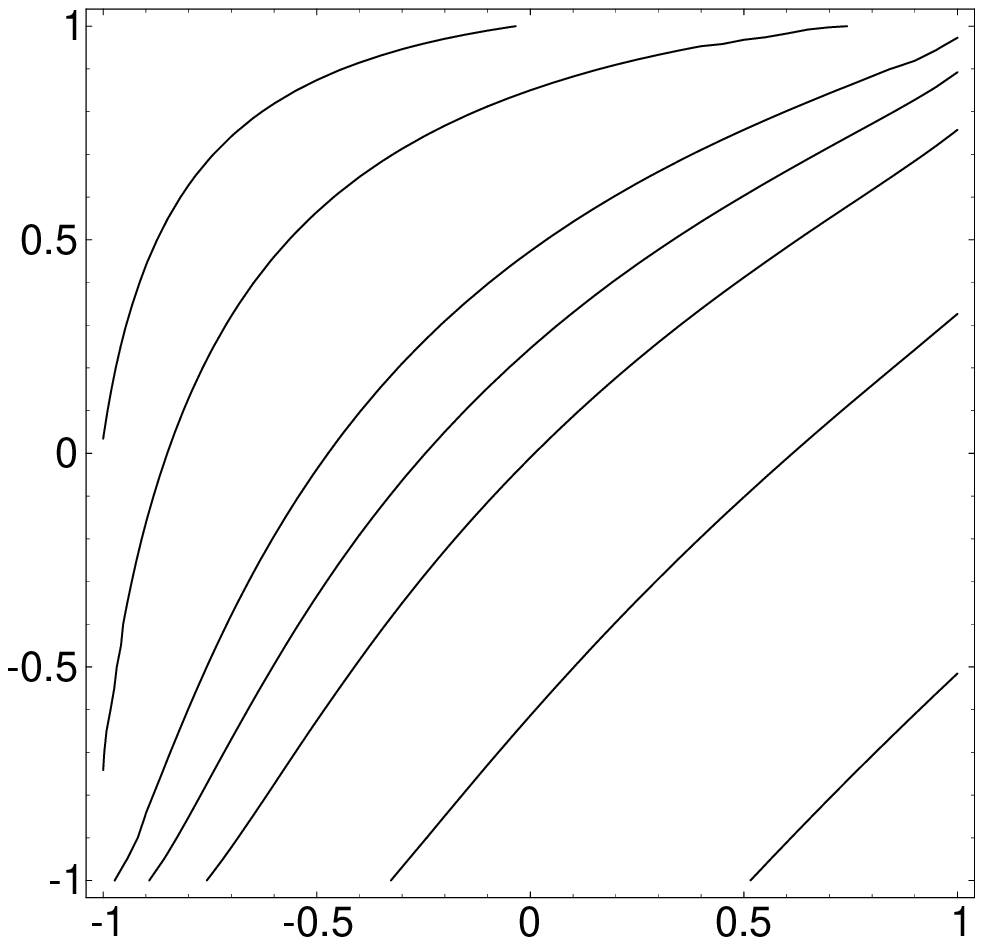}}
	\put(3.,7.5){\fbox{ $S=|{\mathcal A}_{\rm T}|\sqrt{{\mathcal L}\cdot \sigma}$}}
		\put(6.5,-.3){$P_{e^-} $}
		\put(0,7.3){$P_{e^+} $}
	\put(1.3,6.1){\footnotesize 4}
	\put(2.,5.5){\footnotesize 2}
	\put(2.9,4.6){\footnotesize 2}
	\put(3.6,3.8){\footnotesize 4}
	\put(4.1,3.4){\footnotesize 6}
	\put(4.8,2.5){\footnotesize 10}
	\put(6.4,1.2){\footnotesize 15}
 \end{picture}
\vspace*{.1cm}
\caption{
Contour lines of the standard deviations  $S$
in the $P_{e^-}$--$P_{e^+}$ plane
for $M_2=400$ GeV,  $|\mu|=240$ GeV, $\varphi_{M_1}=0.1\pi $ 
and $\varphi_{\mu}=0$, taking
$\tan \beta=10$, $m_0=100$ GeV, 
$\sqrt{s}=500$ GeV
and ${\mathcal L}=500~{\rm fb}^{-1}$.
\label{plot_4}}
\end{minipage}
\end{figure}

In Fig.~\ref{plot_3} we show the 
$\varphi_{\mu}$--$\varphi_{M_1}$ dependence of 
${\mathcal A}_{\rm T}$ for 
$|\mu|=240$ GeV and $M_2=400$ GeV. The value
of ${\mathcal A}_{\rm T}$ depends stronger on $\varphi_{M_1}$
than on $\varphi_{\mu}$. It is remarkable,
that maximal phases $\varphi_{\mu},\varphi_{M_1}=\pm \pi/2$ do
not  lead to the highest values of ${\mathcal A}_{\rm T}$.

The relative statistical error of
${\mathcal A}_{\rm T}$ is given by $\delta {\mathcal A}_{\rm T} = 
\Delta {\mathcal A}_{\rm T}/|{\mathcal A}_{\rm T}| = 
S/(|{\mathcal A}_{\rm T}| \sqrt{N})$ \cite{olaf},
with $S$ standard deviations
and $N={\mathcal L} \cdot\sigma$ the number of events with 
${\mathcal L}$ the total integrated luminosity and 
$\sigma$ the total cross section. Assuming $\delta {\mathcal A}_{\rm T}
\approx1$, it follows $S \approx |{\mathcal A}_{\rm T}| \sqrt{N}$.
For measuring ${\mathcal A}_{\rm T}$, it is thus crucial to have
large ${\mathcal A}_{\rm T}$ and large $\sigma$, which
can both be enhanced by using longitudinally polarized $e^+$ and $e^-$ 
beams.
In Fig.~\ref{plot_4} we show the contour lines of
the standard   deviations $S =|{\mathcal A}_{\rm T}| 
\sqrt{{\mathcal L}\cdot\sigma}$ for  
${\mathcal L}= 500$ fb$^{-1}$. 
In this scenario, 
S can be enhanced for positive $P_{e^-}$ and negative $P_{e^+}$.

%

\subsection{Asymmetry ${\mathcal A}_{\rm CP}$
	\label{asymmetryACP}}

In Fig.~\ref{plot_5}a we show the 
$\varphi_{A_{\tau}}$--$\varphi_{M_1}$ dependence 
of the $\tau$ polarization asymmetry ${\mathcal A}_{\rm CP}$, 
Eq.~(\ref{ACP}), for $\varphi_{\mu}=0$, $|\mu|=300$ GeV and
$M_2=200$ GeV.
We have chosen a small value of $\tan\beta=5$ and a large value of 
$|A_{\tau}|=1$ TeV because ${\mathcal A}_{\rm CP}$ increases with  
increasing $|A_{\tau}| \gg |\mu|\tan\beta$ \cite{staupol}.
The cross section $\sigma=
\sigma(e^+e^-\to\tilde\chi^0_1\tilde\chi^0_2 ) \times
{\rm BR}(\tilde \chi^0_2\to\tilde\tau_1^+\tau^-)$ is shown
in Fig.~\ref{plot_5}b.
Also  $\sigma$ is very sensitive to variations of the 
phases and varies between 30 fb and 180 fb.
The choice of $(P_{e^-},P_{e^+})=(-0.8,0.6)$ enhances
$\sigma$ and ${\mathcal A}_{\rm CP}$.

The feasibility for measuring ${\mathcal A}_{\rm CP}$
depends strongly on the sensitivity $S$ for measuring 
the $\tau$ polarization \cite{davier}. In \cite{atwood}
a sensitivity of $S=0.35$ has been obtained.
For an ideal detector and  considering  
statistical errors only, the sensitivity for measuring 
${\mathcal A}_{\rm CP}$ at $95\%$ C.L.  can be calculated to 
$S=\sqrt{2}/(|{\mathcal A}_{\rm CP}|\sqrt{N})$ \cite{staupol}.
In Fig.~\ref{plot_6} we show $S$ 
in the $P_{e^-}$--$P_{e^+}$ plane.
A sensitivity of $S<0.35$ can be obtained using polarized
beams with negative $P_{e^-}$ and positive $P_{e^+}$.

In Fig.~\ref{plot_7} we show 
for $M_2=400$ GeV and  $|\mu|=300$ GeV
the contour lines of ${\mathcal A}_{\rm CP}$
in the $\varphi_{\mu}$--$\varphi_{M_1}$ plane.
The asymmetry ${\mathcal A}_{\rm CP}$ is very sensitive to
variations of the phases $\varphi_{M_1}$ and $\varphi_{\mu}$,
and can reach values up to 65\%.

%
\FIGURE[h]{
\setlength{\unitlength}{1cm}
\begin{picture}(10,8)(2.65,0)
   \put(0,0){\includegraphics{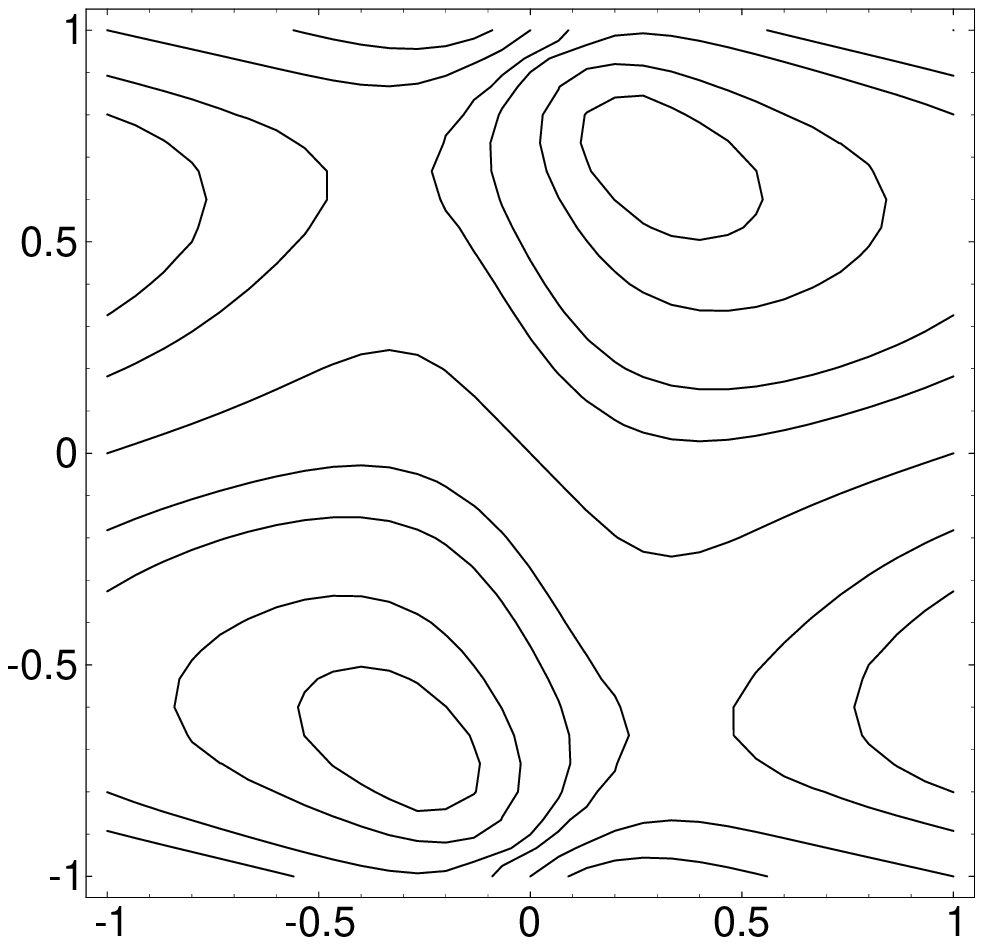}}
	\put(3.,7.4){\fbox{${\mathcal A}_{\rm CP}$ in \% }}
	\put(6.,-0.3){$ \varphi_{A_{\tau}}/\pi$ }
	\put(0,7.4){$ \varphi_{M_1}/\pi$ }
	\put(6.5,6.6){\scriptsize$-15$}
	\put(4.8,5.5){\scriptsize$-50$}
	\put(5.3,4.9){\scriptsize$-40$}
	\put(5.7,4.45){\scriptsize$-25$}
	\put(6.,3.8){\scriptsize$-15$}
	\put(5.,3.1){\scriptsize$0 $}
		\put(1.,5.4){\scriptsize$-25$}
		\put(1.9,5.5){\scriptsize$-15$}
		\put(2.7,6.1){\scriptsize$0$}
		\put(2.8,6.7){\scriptsize$15$}
	\put(5.5,1.8){\scriptsize$15$}
	\put(6.4,1.9){\scriptsize$25 $}
	\put(4.95,1.2){\scriptsize$0$}
	\put(4.55,.7){\scriptsize$-15$}
			\put(1.,.7){\scriptsize$15$}
			\put(2.4,1.9){\scriptsize$50$}
			\put(1.9,2.3){\scriptsize$40$}
			\put(1.6,2.8){\scriptsize$25$}
			\put(1.2,3.4){\scriptsize$15$}
		\put(2.9,4.1){\scriptsize$0 $}
\put(0.5,-.3){Fig.~\ref{plot_5}a}
	\put(8,0){\includegraphics{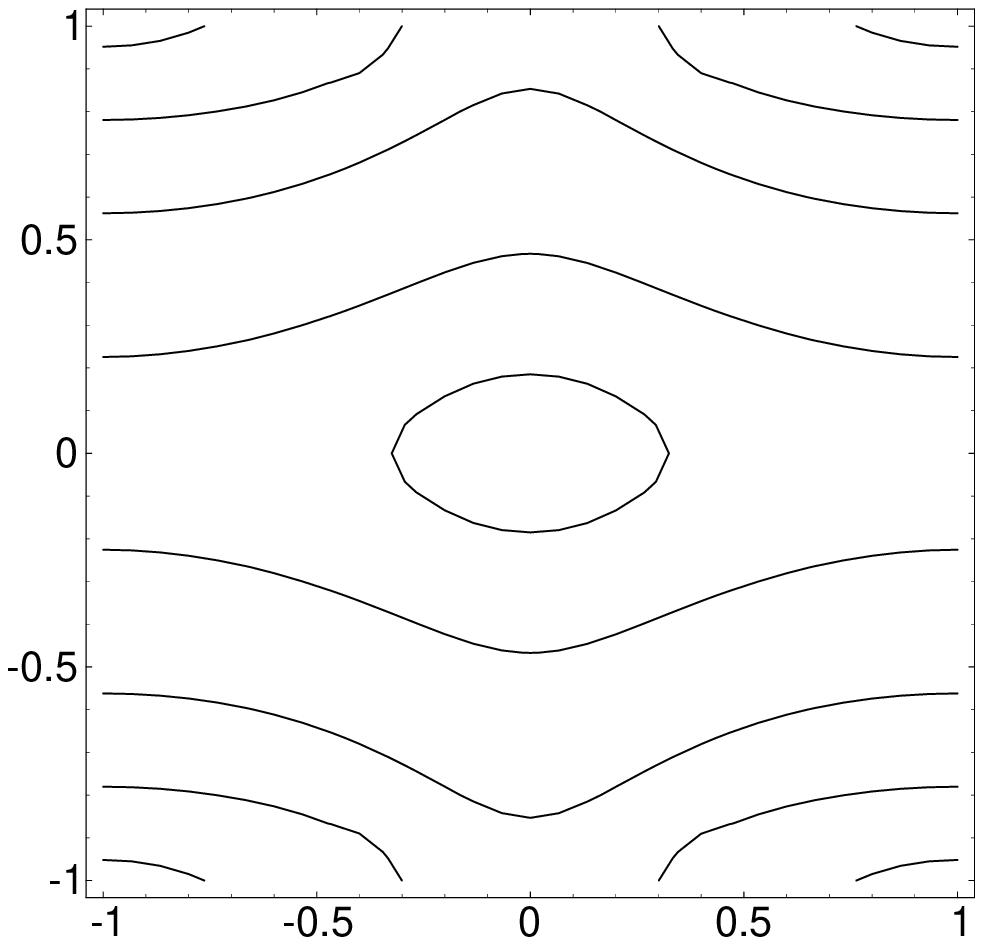}}
	\put(9.5,7.4){\fbox{$\sigma(e^+e^- \to\tilde\chi^0_1 
				\tilde\tau_1^+ \tau^- )$ in fb}}
	\put(14.,-.3){$ \varphi_{A_{\tau}}/\pi$ }
	\put(8,7.4){$ \varphi_{M_1}/\pi$ }
		\put(9.,6.7){\scriptsize$180$}
		\put(9.9,6.4){\footnotesize$150$}
			\put(14.5,6.7){\scriptsize$180$}
			\put(13.6,6.4){\footnotesize$150$}
	\put(11.7,6.0){\footnotesize$100$}
	\put(11.8,4.8){\footnotesize$50$}
	\put(11.8,3.95){\footnotesize$30$}
		\put(11.8,2.5){\footnotesize$50$}
		\put(11.7,1.4){\footnotesize$100$}
	\put(9.9,.9){\footnotesize$150$}
	\put(8.9,0.6){\scriptsize$180$}
			\put(13.6,.9){\footnotesize$150$}
			\put(14.5,0.6){\scriptsize$180$}
	\put(8.5,-.3){Fig.~\ref{plot_5}b}
\end{picture}
\vspace*{.5cm}
\caption{
	Contour lines of 
	the asymmetry ${\mathcal A}_{\rm CP}$ (\ref{plot_5}a) and 
	$\sigma=\sigma(e^+e^-\to\tilde\chi^0_1\tilde\chi^0_2 ) \times
	{\rm BR}(\tilde \chi^0_2\to\tilde\tau_1^+\tau^-)$ (\ref{plot_5}b)
	in the $\varphi_{A_{\tau}}$--$\varphi_{M_1}$ 
	plane for $\varphi_{\mu}=0$, $|A_{\tau}|=1$ TeV,
	$M_2=200$ GeV,  $|\mu|=300$ GeV,
	taking  $\tan \beta=5$, $m_0=100$ GeV,
	$\sqrt{s}=500$ GeV and $(P_{e^-},P_{e^+})=(-0.8,0.6)$.
\label{plot_5}}
}

%
\begin{figure}[h]
\setlength{\unitlength}{1cm}
\begin{minipage}{0.45\textwidth}
 \begin{picture}(10,7.5)(.4,.5)
	\put(0,0){\includegraphics{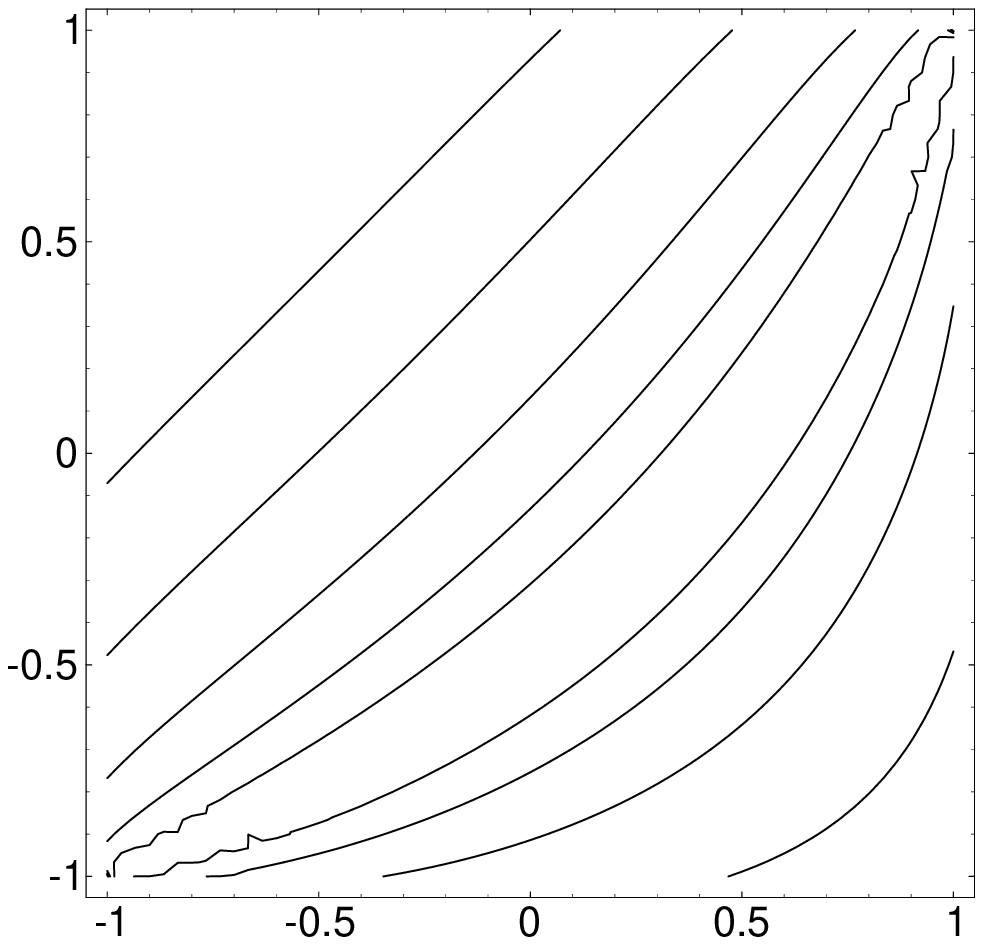}}
	\put(2,7.5){\fbox{ $S=\sqrt{2}/(|{\mathcal A}_{\rm CP}|
       \sqrt{{\mathcal L}\cdot \sigma})$ }}
	\put(6.5,-.3){$P_{e^-}$}
	\put(0.3,7.3){$ P_{e^+}$ }
	\put(2.,5.4){\footnotesize 0.15}
	\put(2.8,4.6){\footnotesize 0.2}
	\put(3.4,4.1){\footnotesize 0.3}
	\put(3.8,3.6){\scriptsize 0.5}
	\put(4.45,3.0){\footnotesize 1}
	\put(4.75,2.65){\footnotesize 1}
	\put(5.3,2.3){\scriptsize 0.5}
	\put(5.7,1.8){\footnotesize 0.3}
	\put(6.4,1.1){\footnotesize 0.2}
 \end{picture}
\vspace*{.5cm}
\caption{Contour lines of the sensitivity $S$
in the $P_{e^-}$--$P_{e^+}$ plane
for $M_2=200$ GeV,  $|\mu|=300$ GeV, 
$\varphi_{A_{\tau}}=0.2\pi$, $|A_{\tau}|=1$ TeV,
$\varphi_{\mu}=\varphi_{M_1}=0$, taking
$\tan \beta=5$, $m_0=100$ GeV, $\sqrt{s}=500$ GeV
and ${\mathcal L}=500~{\rm fb}^{-1}$.
\label{plot_6}}
\end{minipage}
\hspace*{0.1cm}
	\begin{minipage}{0.45\textwidth}
 \begin{picture}(0,8)(.4,0)
	\put(0,0){\includegraphics{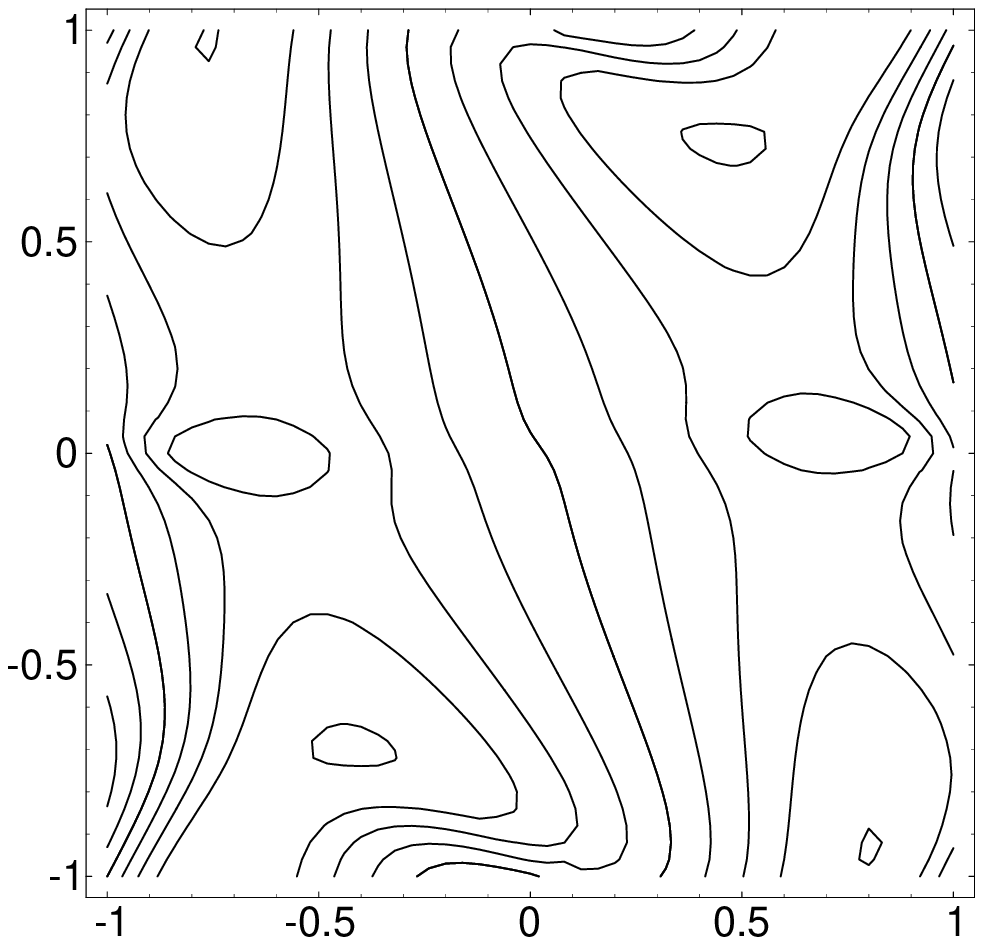}}
	\put(3.,7.5){\fbox{ ${\mathcal A}_{\rm CP} $ in \%}}
		\put(6.5,-.3){$ \varphi_{\mu}/\pi$}
		\put(0,7.3){$ \varphi_{M_1}/\pi$}
	\put(-7.85,0.05){\begin{picture}(1,1)(0,0)
			\setlength{\unitlength}{3.5cm}
	\put(2.96,.45){\scriptsize{65}}
	\put(2.90,.66){\scriptsize{45}}
	\put(2.77,.80){\scriptsize{30}}
	\put(2.54,.85){\scriptsize{0}}
	\put(2.75,1.03){\scriptsize{45}}
	\put(2.90,1.30){\scriptsize{30}}
	\put(2.51,1.35){\scriptsize{15}}
	\put(2.65,1.53){\scriptsize{45}}
	\put(3.23,.80){\scriptsize{15}}
	\put(3.40,.90){\scriptsize{0}}  
\put(3.50,1.00){\scriptsize{-15}}
	\put(4.03,.30){\scriptsize{-65}}
	\put(3.95,.58){\scriptsize{-45}}
	\put(3.75,.83){\scriptsize{-30}}
	\put(3.93,1.07){\scriptsize{-45}}
	\put(3.91,1.25){\scriptsize{-30}}
	\put(4.05,1.35){\scriptsize{-15}}
	\put(4.20,1.25){\scriptsize{0}}
	\put(3.73,1.46){\scriptsize{-45}}
	\put(3.71,1.67){\scriptsize{-65}}
		\setlength{\unitlength}{1cm}
	\end{picture}}
 \end{picture}
\vspace*{.5cm}
\caption{
Contour lines of ${\mathcal A}_{\rm CP} $
in the $\varphi_{\mu}$--$\varphi_{M_1}$ plane
for $M_2=400$ GeV,  $|\mu|=300$ GeV, taking
$\tan \beta=5$, $m_0=100$ GeV, 
$\varphi_{A_{\tau}}=0$, $|A_{\tau}|=250$ GeV,
$\sqrt{s}=500$ GeV and $(P_{e^-},P_{e^+})=(-0.8,0.6)$.
\label{plot_7}}
\end{minipage}
\end{figure}

\section{Summary and conclusion
	\label{Summary and conclusion}}

We have proposed a T and a CP-odd asymmetry in 
$e^+e^- \to\tilde\chi^0_1  \tilde\chi^0_2$
and the subsequent leptonic two-body decay of $\tilde\chi^0_2$,
in order to analyze  CP violation caused by the phases 
$\varphi_{M_1}$, $\varphi_{\mu}$ and $\varphi_{A_{\tau}}$.
In a numerical study for the decay 
$\tilde\chi^0_2 \to \tilde\ell_R \ell_1$,
$\tilde\ell_R \to \tilde\chi^0_1 \ell_2$ with $ \ell_{1,2}= e,\mu$,
we have shown that the asymmetry ${\mathcal A}_{{\rm T}}$ of 
the triple product
$ (\vec p_{e^-}\times\vec p_{\ell_2})\cdot \vec p_{\ell_1}$, 
can reach values up to 6\% even for small phases, e.g. 
$\varphi_{M_1}=0.1\pi$ and $\varphi_{\mu}=0$.
The asymmetry  ${\mathcal A}_{\rm T}$ and the cross section
$\sigma$ can be enhanced by longitudinally polarized beams.
For the neutralino decay 
$\tilde\chi^0_2 \to \tilde\tau_1^{\mp}\tau^{\pm}$,
we have given numerical examples for the 
CP-odd $\tau$ polarization asymmetry ${\mathcal A}_{\rm CP}$,
which is also sensitive to $\varphi_{A_{\tau}}$.
The asymmetry can reach values up to 60\%.
Longitudinally polarized beams enhance ${\mathcal A}_{\rm CP}$
and the cross section, such that the sensitivity for
measuring the $\tau$ polarization can be reduced significantly.
In a numerical example, we have shown that a sensitivity
of $S<0.35$ can be achieved to measure a phase of  
$\varphi_{A_{\tau}}=0.2\pi$.
Depending on the MSSM scenario, the asymmetries should be
accessible in future electron-positron linear collider
experiments with longitudinally polarized beams in 
the 500 GeV c.m.s. energy range.

\acknowledgments

This work was supported by the 'Deutsche Forschungsgemeinschaft'
(DFG) under contract Fr 1064/5-1, by Spanish grants BFM2002-00345,
by the EU network Programme HPRN-CT-2000-00148 and by the EU
Research Training Site contract HPMT-2000-00124.


\begin{thebibliography}{99}


\bibitem{BGSM}
F.~Csikor et al., Phys.\ Rev.\ Lett. {\bf 82} (1999) 21.

\bibitem{Masiero:xj}
For a recent review see, e.g., A.~Masiero and O.~Vives,
New J.\ Phys.\  {\bf 4} (2002) 4.

\bibitem{edmstheo}
For a review see, e.g., P. Nath, Talk at the 9th International 
Conference on Supersymmetry 
and Unification of Fundamental Interactions, 11-17 June 2001, Dubna 
[arXiv:hep-ph/0107325] and references therein.

\bibitem{Barger:2001nu}
see e.g. V.~D.~Barger, T.~Falk, T.~Han, J.~Jiang, T.~Li and T.~Plehn,
Phys.\ Rev.\ D {\bf 64} (2001) 056007
[arXiv:hep-ph/0101106].
%

\bibitem{TDR} TESLA Technical Design Report, Part III, \emph{Physics at an
$e^+e^-$ Linear Collider}, eds. R.-D.~Heuer, D.~Miller, F.~Richard and
P.~Zerwas, [arXiv:hep-ph/0106315].

\bibitem{donoghue} J.~F.~Donoghue,
 Phys. Rev. {\bf D 18} (1978) 1632;
G.~Valencia,
[arXiv:hep-ph/9411441];
G.C.~Branco, L.~Lavoura, and J.P.~Silva, {\em CP violation}, 
Oxford University Press, New York, 1999.

\bibitem{oshimo} Y.~Kizukuri and N.~Oshimo,
                 Phys. Lett. {\bf B 249} (1990) 449.

\bibitem{choi1} S.~Y.~Choi, H.~S.~Song and W.~Y.~Song,
						  Phys. Rev. {\bf D 61} (2000) 075004.
					
\bibitem{kali}
S.~Y.~Choi, J.~Kalinowski, G.~Moortgat-Pick and P.~M.~Zerwas,
Eur. Phys. J. {\bf C 22} (2001) 563; 
Addendum-ibid. {\bf C 23} (2001) 769.
						  
\bibitem{gudi1} 
G.~Moortgat-Pick, H.~Fraas, A.~Bartl and W.~Majerotto, 
Eur. Phys. J. {\bf C 9} (1999) 521;
Erratum-ibid. {\bf C 9} (1999) 549.
	  

\bibitem{olaf} A.~Bartl, H.~Fraas, O.~Kittel and W.~Majerotto,
[arXiv:hep-ph/0308141]; 
A.~Bartl, H.~Fraas, O.~Kittel and W.~Majerotto, 
[arXiv:hep-ph/0308143].

\bibitem{karl}
K. Hohenwarter-Sodek, Diploma thesis, University of Vienna (2003), in
german.

\bibitem{Bartl:2003ck}
A.~Bartl, H.~Fraas, T.~Kernreiter and O.~Kittel,
arXiv:hep-ph/0306304.

\bibitem{Renard}
F.~M.~Renard, \it Basics of Electron Positron Collisions, 
\rm Editiors Frontieres, 
Dreux (1981).

\bibitem{staupol} A.~Bartl, T.~Kernreiter, O.~Kittel,
[arXiv:hep-ph/0309340].

\bibitem{Choi:2003pq}
S.~Y.~Choi, M.~Drees, B.~Gaissmaier and J.~Song,
arXiv:hep-ph/0310284.

\bibitem{POWER} A.~Bartl, H.~Fraas, T~Kernreiter, 
O.~Kittel and W.~Majerotto, [arXiv:hep-ph/0310011].

\bibitem{gudi2}
G.~Moortgat-Pick,
in {\it Proc. of the APS/DPF/DPB Summer Study on the Future of Particle Physics (Snowmass 2001) } ed. N.~Graf,
eConf {\bf C010630}, E3008 (2001)
[arXiv:hep-ph/0202082];
G.~Moortgat-Pick, A.~Bartl, H.~Fraas and W.~Majerotto,
Eur.\ Phys.\ J.\ C {\bf 18} (2000) 379.

\bibitem{hall} L.~J.~Hall and J.~Polchinski,
Phys. Lett. {\bf B 152} (1989) 335.

\bibitem{davier}
M.~Davier, L.~Duflot, F.~Le Diberder and A.~Rouge,
Phys.\ Lett.\ B {\bf 306} (1993) 411.

\bibitem{atwood} D. Atwood, G. Eilam and A. Soni,
         Phys. Rev. Lett. {\bf 71} (1993) 492.


\end{thebibliography}
\end{document}